\documentclass{article}
\usepackage[dvips]{graphicx}

\begin{document}
\begin{center}
\centerline{\large \bf On the physical nature}
\centerline{\large \bf of Anderson localization effect} 
\end{center}

\vspace{3 pt}
\centerline{\sl V.A. Kuz'menko\footnote{Electronic 
address: kuzmenko@triniti.ru}}

\vspace{5 pt}

\centerline{\small \it Troitsk Institute for Innovation and Fusion 
Research,}
\centerline{\small \it Troitsk, Moscow region, 142190, Russian 
Federation.}

\vspace{5 pt}

\begin{abstract}

An alternative explanation of the physical nature of Anderson 
localization phenomenon and one of the most direct ways of its 
experimental study are discussed.

\vspace{12 pt}
\end{abstract}

\vspace{8 pt}

Originally Anderson localization effect was found for transport of 
electrons in a crystal lattice [1]. In recent years it has been extensively 
studied for more convenient object - scattering of light, ultrasound, 
microwaves [2-5]. The essence of the effect is that the transport of the 
scattered light is not consistent with the diffusion model and needs to 
introduce the position-dependent diffusion coefficient [6].  

For many years the established common opinion about physical nature of 
Anderson localization effect supposes that the main role here 
plays the interference of the scattered waves. It is supposed that Anderson 
localization "... originates from constructive interference 
of waves traveling in loop trajectories - pairs of time-reversed 
paths returning to the same point. ... a wave may return to a position 
it has previously visited after a random walk, and there is always the 
time-reversed path which yields identical phase delay. Constructive 
interference of the waves from the reversed loops increases wave 
(energy) density at the original position and decreases the flux, 
giving the so-called weak localization effect. This is the basic mechanism 
for the suppression of wave diffusion, which eventually 
leads to Anderson localization" [6]. However, this quite fantastical 
explanation can not solve the main problem here: why the scattered photons 
(waves) return to the same initial point. 

We shall discuss here the physical explanation which does not need in any 
interference phenomena [7]. We believe that Anderson localization of 
light is quite simple example of numerous manifestations of a fundamental 
property of quantum physics: its time reversal noninvariance or inequality 
of forward and reversed processes [8, 9]. It is interesting, that this 
concept, in fact, was proved already, although it is not recognized as yet. 
The essence of this concept is that the cross-section of reversed transition 
(in contrast to those for forward transition) has very sharp dependence 
from the number of physical parameters. Its cross-sections can and really 
differ in many orders of magnitude (although its integral cross-sections 
are equal). In Fig.1 
we show the supposed dependences of cross-sections of forward and 
reversed optical transitions in two-level system from such parameters as the 
frequency and phase of laser radiation, orientation of molecule in space, 
the phase of atom vibration in molecule and even the position of the atom 
in space. As a result we have following supposed hierarchy of the discussed 
cross-sections:

\begin{equation}
\sigma_{FR} \gg \sigma_{PR} \gg \sigma_{F} > \sigma_{B} 
\end{equation}

Let us discuss the most common case of light scattering on the free atoms. 
Scattering of the photon in a sideway direction is a case of forward 
($\sigma_{F} $) or backward ($\sigma_{B} $) transition 
(Fig. 2a). In this case the direction of movement and position in 
space for photon and atom are changed. Scattering of photon in the backward 
direction (Fig. 2b) corresponds to partially reversed 
transition ($\sigma_{PR} $). In this case the photon can returns to 
the initial point, but its direction of movement is changed. The 
atom receives two photon recoil moments and can change its position 
and direction of movement (not shown in Fig. 2). Scattering of 
photon in the forward direction (Fig. 2c) is the most close to a 
fully reversed transition ($\sigma_{FR} $). In this case the 
direction of photon's movement and position of atom remain unchanged. 
But the position of the photon in space is changed. We can use the 
two massive mirrors (Fig. 2d), which will allow the photon (and the quantum 
system as a whole) to return exactly to its initial state. This case 
corresponds to fully reversed process. The mirror can also improve the 
situation with backscattering (Fig. 2e). This variant 
was widely known early as the wave front reversing (conjugation) 
[10-14]. So, we can expect following hierarchy of the discussed 
cross-sections for the processes of Fig.2:

\begin{equation}
\sigma_{d}\ > \sigma_{c}\ ; \sigma_{e}\ > \sigma_{b}\ > \sigma_{a} 
\end{equation}

It is worth to note that in variants d and e in Fig. 2 we deal with the 
problem of nonlocality: how are the photon and atom aware about the 
existence of mirrors? This is the same problem as the classical problem 
in quantum physics of diffraction from the two slits. In Bohm's theory [15] 
this problem is solved by introducing the so-called non-local quantum 
potential, which may be considered as equivalent to the memory 
of a quantum system (as a whole) about the initial state [16]. 

It is clear from the present explanation that for study the effect of 
Anderson localization (and many other phenomena in quantum physics), we 
need to measure and compare the differential cross sections of forward, 
reversed and partially reversed transitions. The most convenient object 
today for experimental study of differential cross-sections of quantum 
transitions is the so-called Bloch oscillations of cold atoms in vertical 
optical lattice [17, 18]. This is the same variant as in Fig. 2d but with 
vertical resonator. Here the cold atoms under action of gravity freely fall 
down in vacuum. In certain point the specific scattering of photons takes 
place: one upward photon is absorbed and one downward photon is emitted. 
As a result, the recoil momentum returns the atoms exactly into the initial 
point (state). 

This is the most clean example of fully reversed quantum transition today. 
This transition has highest possible differential cross-section. All other 
possible scattering processes will be partially reversed or forward and will 
have lesser differential cross-sections (Fig.1). The authors in 
[18] had observed nearly ${10^4}$ of Bloch oscillations of cold atoms 
in vertical optical lattice. It means that differential cross-section 
of fully reversed quantum transition here in more than four orders of 
magnitude exceeds differential cross-sections of other possible scattering 
processes, which can destroy the Bloch oscillations. The dependences of the 
differential cross-sections from physical parameters (frequency and phase 
of laser radiation, its direction, the position of the atom in space) 
nobody experimentally studied till now. This is not a very difficult task. 
However, the main problem here is that before such experiments our 
physicists must reject their paradigm:"... a remarkable fundamental fact of 
nature: all known laws of physics are invariant under time reversal" [19].

\begin{figure}
\newpage
\begin{center}
\includegraphics[scale=1.0]{Figure1.ps}
\end{center}
\end{figure}

\begin{figure}
\newpage
\begin{center}
\includegraphics[scale=1.0]{Figure2.ps}
\end{center}
\end{figure}

\end{document}